\documentclass[aps,twocolumn,superscriptaddress,showpacs,floatfix]{revtex4}%
\usepackage{amsfonts}
\usepackage{amsmath}
\usepackage{amssymb}
\usepackage{graphicx}
\usepackage{graphicx}%
\begin{document}
\title{Transformation of a single photon field into bunches of pulses}
\author{R.N. Shakhmuratov}
\affiliation{Kazan Physical-Technical Institute, Russian Academy of Sciences,
10/7 Sibirsky Trakt, Kazan 420029 Russia}
\affiliation{Kazan Federal University, 18 Kremlyovskaya Street, Kazan 420008 Russia}
\author{F.G. Vagizov}
\affiliation{Kazan Physical-Technical Institute, Russian Academy of Sciences,
10/7 Sibirsky Trakt, Kazan 420029 Russia}
\affiliation{Kazan Federal University, 18 Kremlyovskaya Street, Kazan 420008 Russia}
\affiliation{Department of Physics and Institute for Quantum Studies, TAMU, College
Station, Texas 77843-4242, USA}
\author{V.A. Antonov}
\affiliation{Kazan Federal University, 18 Kremlyovskaya Street, Kazan 420008 Russia}
\affiliation{Institute of Applied Physics of the Russian Academy of Sciences, Nizhny
Novgorod 603950, Russia}
\affiliation{N.I. Lobachevsky State University of Nizhny Novgorod,
Nizhny Novgorod, 603950, Russia}
\author{Y.V. Radeonychev}
\affiliation{Institute of Applied Physics of the Russian Academy of Sciences, Nizhny
Novgorod 603950, Russia}
\affiliation{N.I. Lobachevsky State University of Nizhny Novgorod,
Nizhny Novgorod, 603950, Russia}
\author{O. Kocharovskaya}
\affiliation{Department of Physics and Institute for Quantum Studies, TAMU, College
Station, Texas 77843-4242, USA}
\date{{ \today}}

\begin{abstract}
We propose a method to transform a single photon field into bunches of pulses
with controllable timing and number of pulses in a bunch. This method is based
on transmission of a photon through an optically thick single-line
absorber vibrated with a frequency appreciably exceeding the width of
the absorption line. The spectrum of the quasi-monochromatic incoming photon
is 'seen' by the vibrated absorber as a comb of equidistant spectral components
separated by the vibration frequency. Tuning the absorber in resonance with
$m$-th spectral component transforms the output radiation into bunches of pulses
with $m$ pulses in each bunch. We experimentally demonstrated the proposed technique
with single 14.4 keV photons and produced for the first time gamma-photonic
time-bin qudits with dimension $d=2m$.
\end{abstract}
\maketitle

\textbf{Introduction}. Rapid development of quantum information technologies
demands to invent and bring into routine new methods to control and process
single photons. In vast majority of quantum protocols photon polarization is
used as the information carrier, see, for example, review \cite{Gisin2002R}.
Time-bin qubits, proposed and implemented in Refs. \cite{Gisin1999,Gisin2002},
were practically the first examples how time domain can be involved into the
information coding by splitting a single photon into two pulses with a fixed
phase difference. The information carried by such a photon is well protected
during its propagation in optical fiber since cross talk, usually influencing
polarization states of a photon, is excluded. Information coding
\cite{Gisin1999,Gisin2002} is implemented by the interferometer having different
lengths of the arms and a phase shifter placed in the long arm of the
interferometer. Unbalanced three path interferometers were used in
\cite{Gisin2004} to create qutrits. Recently, new methods to process time-bin
qubits \cite{Metcalf,Donohue} and to create time-bin qubits, qutrits, and
ququads \cite{Kuhn} were reported.

In this paper we propose to split a photon in time bins by transmission through
a thick resonant absorber experiencing phase modulation of its interaction with
the field. We test our proposal in gamma domain where a single photon is
produced by a naturally decaying nucleus. Radioactive $^{57}$Co is ideal for
test experiments since coherence length of produced photons is incredibly
long and falls into domain where simple electronics may be applied to
treat experimental data. The source containing these nuclei is commercially
available and relatively cheap. Moreover, the detectors in gamma domain
have high quantum efficiency and almost no dark counts. Our method
to control the shape of a single-photon wave packet is applicable to all spectral
domains, including optical domain where atoms may play a role of optical memory
and processors for quantum information. The modulation of the atom-field
interaction in optical domain can be realized by Zeeman/Stark effects, as it
was discussed in Refs. \cite{Rad2010,Rad2013} in connection with a proposal
of the new technique to produce extremely short pulses, or by phase modulators.
The method is flexible to make fine tuning the amplitudes,
time, and intervals between pulses produced from a single photon.

We have to mention the currently available experimental techniques, developed
for the single gamma-photon waveform shaping, which could be also applied to
extend the methods of quantum information processing. They include magnetic
switching \cite{Shvydko91,Shvydko96,Palffy09,Palffy12}, modulation technique
\cite{Smirnov,Vagizov}, and step-wise phase modulation of the radiation field
\cite{Helisto91,Helisto93,Helisto84,Ikonen,vanBurk,Shakhmuratov11,Shakhmuratov13}.

\textbf{Basic idea}. Single photon, emitted by a source nucleus, has a
Lorenzian spectrum whose width is mostly defined by the lifetime of the
excited state nucleus. We transmit such a photon through a thick resonant
absorber with a single resonance line having approximately the same width as
the spectrum of the radiation field. The absorber experiences periodic
mechanical vibrations along the photon propagation direction. They are induced
by the piezoelectric transducer. In the reference frame of the piston-like
vibrating absorber the field phase oscillates as $(2\pi \delta/\lambda) \sin\Omega
t$, where $\Omega$ and $\delta$ are the frequency and amplitude of the periodical
displacements, $\lambda$ is the wavelength of the radiation field. The
probability amplitude of the radiation field in the laboratory reference
frame, $a_{L}(t,z)=a_{0}(t)\exp(-i\omega_{s}t+ik_{s}z)$, is transformed to
$a_{A}(t,z)=a(t,z)_{L}\exp[ip\sin(\Omega t+\varphi)]$ in the vibrating
absorber reference frame. Here $\omega_{s}$, $k_{s}$ are the frequency and
wave number of the radiation field, $z$ is a coordinate on the axis
$\mathbf{z}$, along which the photon propagates, $p=2\pi \delta/\lambda$ is the
modulation index, and $\varphi$ is the relative phase of the absorber
vibration. The probability amplitude $a_{A}(t,z)$ can be expressed as
\begin{equation}
a_{A}(t,z)=a_{L}(t,z)%
%TCIMACRO{\dsum \limits_{m=-\infty}^{+\infty}}%
%BeginExpansion
{\displaystyle\sum\limits_{m=-\infty}^{+\infty}}
%EndExpansion
J_{m}(p)e^{im(\Omega t+\varphi)}, \label{Eq1}%
\end{equation}
where $J_{m}(p)$ is the Bessel function of the $m$-th order. From this
expression it is obvious that the vibrating absorber 'sees' the incident
radiation field as an equidistant frequency comb with spectral components
$\omega_{s}-m\Omega$ having Lonrenzian shape each and intensities, which are
proportional to $J_{m}^{2}(p)$. Let us say that the $m$-th component of this
field is tuned in resonance with the absorber. If the halfwidth of the
components $\gamma_{s}$ (it is defined by the halfwidth of the spectrum of the
incident field, $a_{L}(t,z)$) is much smaller than the distance between
neighboring components, $\Omega$, then we may assume that only resonant
component interacts with the absorber and others pass through without
interaction. Within the adopted approximation only the interacting component
is coherently scattered by nuclei of the absorber. For week fields
the output radiation field can be expressed as a sum of the incident
and coherently scattered radiation fields \cite{Shakhmuratov12}. In case
of a single-line resonant radiation these fields interfere destructively,
resulting in the radiation damping. For the frequency comb only the
resonant component is coherently scattered by resonant nuclei and
the scattered field interferes with the whole frequency comb at the exit
of the absorber. Therefore the output radiation field reveals unusual
properties.

The shape of a single photon wave-packet is described by the exponentially
decaying function with the rate $\gamma_{s}$,
\begin{equation}
a_{L}(t-t_{0})=\Theta(t-t_{0})\exp[-(i\omega_{s}+\gamma_{s})(t-t_{0}%
)+ik_{s}z], \label{Eq2}%
\end{equation}
where $\Theta(t-t_{0})$ is the Heaviside step function and $t_{0}$ is the
moment of time when the excited state of the source nucleus is formed. Such a
shape is typical for single photon wave-packets if we know the time of
formation of the excited state particle producing this single photon (see, for
example Refs. \cite{Lynch,Lounis}). In our case we also know this time to be
able to reconstruct the photon shape and its following transformation by the
absorber. In our experiments the source nucleus, $^{57}$Co, decays by electron
capture to $^{57m}$Fe, which decays in turn by emission of a 122keV photon,
followed by a 14.4 keV photon to the ground state $^{57}$Fe. In this cascade
decay the detection of 122 keV photon heralds the formation of 14.4 keV
excited state of $^{57}$Fe nucleus in the source (see, for example, Ref.
\cite{Lynch}). The absorber contains ground state $^{57}$Fe nuclei, which are
resonant for 14.4 keV photons.

The propagation of the field, Eq. (\ref{Eq2}), through a single-line resonant
absorber (not vibrating) can be described classically, Ref. \cite{Lynch}, or
quantum mechanically, Ref. \cite{Harris61}. The result is well known in gamma
domain \cite{Lynch,Harris61} and in quantum optics \cite{Crisp}. In the
simplest case if $\omega_{s}=\omega_{a}$ and $\gamma_{s}=\gamma_{a}$, where
$\omega_{a}$ and $\gamma_{a}$ are the resonant frequency and halfwidth of the
absorption line of the absorber, respectively, the output probability
amplitude is%
\begin{equation}
a_{out}(t,l)=a_{L}(t,l)J_{0}\left(  2\sqrt{bt}\right)  , \label{Eq3}%
\end{equation}
where $l$ is the physical thickness of the absorber, $t_{0}=0$, and $t$ is the
local time $t-l/c$. Below we disregard this retardation since physical length
of the absorber is small and retardation time $l/c$ is short with respect to
the time scale of the amplitude evolution. The parameter $b=$ $\gamma_{a}%
T_{a}/2$ is defined by the product of the decay rate of the nuclear coherence
$\gamma_{a}$ and optical depth of the absorber, which is $T_{a}=\alpha_{B}l$,
where $\alpha_{B}=N\sigma$ is the Beer's law absorption coefficient applicable
to a monochromatic radiation tuned in resonance, $N$ is the density of $^{57}%
$Fe nuclei in the absorber, and $\sigma$ is the cross section of resonant
absorption for 14.4 keV transition. The inverse value of $b$ is usually
referred to as superradiant time, $T_{SR}=1/b$. Here we disregard recoil
processes in nuclear absorption and emission assuming that recoilless fraction
(Deby-Waller factor) is $f_{a}=1$. These processes will be taken into account
for experimental data treatment.

Since we are interested only in the detection probability of a photon, which
is equivalent to the radiation intensity, the cases when the absorber vibrates
with respect to the source at rest or vice versa give the same result. For
simplicity we consider the case of the vibrating source. Then, the radiation
field at the exit of the absorber at rest is the sum of the comb, Eq.
(\ref{Eq1}), and the coherently scattered field. We suppose that the frequency
component $\omega_{s}-m\Omega$ is in resonance with the absorber. Then,
according to Eq. (\ref{Eq3}) the probability amplitude of the coherently
scattered field is%
\begin{equation}
a_{sct}(t,l)=a_{L}(t,l)S_{m}(t)e^{im(\Omega t+\varphi)},\label{Eq4}%
\end{equation}
where $S_{m}(t)=J_{m}(p)\left[  J_{0}\left(  2\sqrt{bt}\right)  -1\right]  $.
This field is just the output field $a_{out}(t,l)$ for the component
$a_{L}(t,z)J_{m}(p)\exp[im(\Omega t+\varphi)]$ minus this component if it
would propagate without interaction with the absorber
\cite{Shakhmuratov09,Shakhmuratov11}. Simple calculation of the probability
$P(t)=\left\vert a_{L}(t,l)+a_{sct}(t,l)\right\vert ^{2}$ of the output
radiation field gives \begin{figure}[ptb]
\resizebox{0.4\textwidth}{!}{\includegraphics{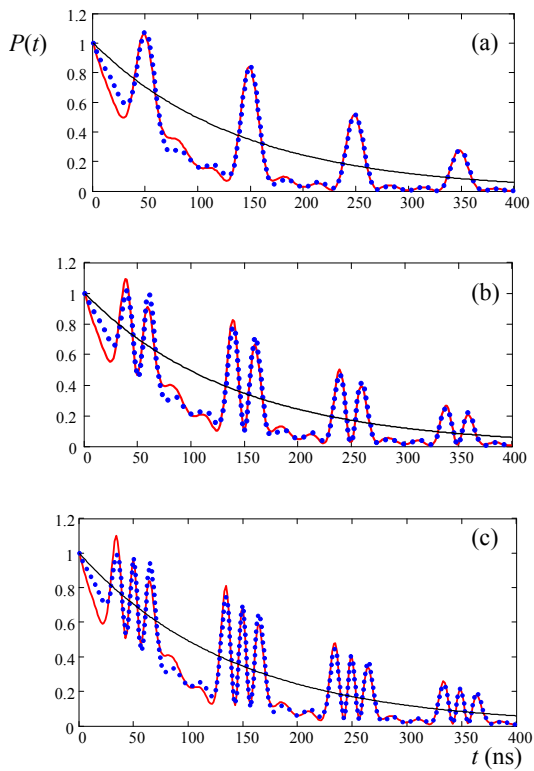}}\caption{(color on
line) Time dependence of the detection probability of a photon, $P(t)$, at the
exit of the absorber vibrating with the frequency $\Omega=10$ MHz and phase
$\varphi=0$. Optical thickness of the absorber is $T=5.2$ and $\gamma
_{a}=\gamma_{s}=1.13$ MHz. The frequency of the radiation field $\omega_{s}$
is tuned in resonance with the first sideband $\omega_{a}+\Omega$ (a), the
second sideband $\omega_{a}+2\Omega$ (b), and the third sideband $\omega
_{a}+3\Omega$ (c). The value of the modulation index is taken optimal in each
case, i.e., $p_{1}=1.8$ (a), $p_{2}=3.1$ (b), and $p_{3}=4.2$ (c), (see the
text). Dotted line (in blue) corresponds to the analytical approximation
(\ref{Eq5}) and solid line (in red) demonstrates the result obtained from the
exact solution (\ref{Eq6}).}%
\label{fig:1}%
\end{figure}%
\begin{equation}
P(t)=\Theta(t)e^{-2\gamma t}\left[  1+2S_{m}(t)\cos\psi_{m}(t)+S_{m}%
^{2}(t)\right]  ,\label{Eq5}%
\end{equation}
where $\gamma=\gamma_{a}=\gamma_{s}$ and $\psi_{m}(t)=m(\Omega t+\varphi
)-p\sin(\Omega t+\varphi)$. Time evolution of the probability $P(t)$ for
$m=1$, $2$, and $3$, is shown in Fig. 1 (a, b, and c, respectively) where the
results of approximate Eq. (\ref{Eq5}) are compared with the exact expression
$P_{ext}(t)=\left\vert a_{ext}\right\vert ^{2}$, which is obtained if we
calculate the probability amplitude without any assumptions (see, for example,
Ref. \cite{Ikonen}), i.e.,%
\begin{equation}
a_{ext}(t,l)=a_{A}(t,l)-b%
%TCIMACRO{\dint \limits_{0}^{t}}%
%BeginExpansion
{\displaystyle\int\limits_{0}^{t}}
%EndExpansion
a_{A}(t-\tau,l)j_{1}(b\tau)e^{-\gamma_{a}\tau-i\omega_{a}\tau}d\tau
,\label{Eq6}%
\end{equation}
where $j_{1}(b\tau)=J_{1}\left(  2\sqrt{b\tau}\right)  /\sqrt{b\tau}$. Small
misfits can be almost excluded if we take into account the contribution of the
two neighboring satellites, both red and blue detuned from the resonant
component of the comb, see Supplemental Material (SM).

We see that the shape of the photon wave packet is transformed into bunches of
pulses with the number of pulses per bunch equal to $m$. The pulses are
produced due to constructive interference of the incident field with
coherently scattered field when $\psi_{m}(t)=(2n+1)\pi$, $n=0,1,2,...$, while
the dark windows appear due to destructive interference when $\psi
_{m}(t)=2n\pi$, if $S_{m}(t)$ is negative in both cases. The probability has
maxima, corresponding to pulses, $[1-S_{m}(t)]^{2}\exp(-2\gamma t)$ and minima
$[1+S_{m}(t)]^{2}\exp(-2\gamma t)$, relevant to the radiation drop. The most
pronounced pulses appear if the intensity of the resonant component [$\sim
J_{m}^{2}(p)$] has global extremum. For example, the maxima, predicted by Eq.
(\ref{Eq5}) without exponential factor $\exp(-2\gamma t)$ and assuming that
$J_{0}\left(  2\sqrt{bt}\right)  \approx0$, which means that the scattered
field had time to fully develop, are $[1+J_{1}(p_{1})]^{2}=2.5$ if $p_{1}%
=1.8$; $[1+J_{2}(p_{2})]^{2}=2.1$ if $p_{2}=3.1$; and $[1+J_{3}(p_{3}%
)]^{2}=2.06$ if $p_{3}=4.2$, i.e., the intensity of the pulses exceeds almost
two times the intensity of the radiation field if it would not interact with
the absorber. The radiation field between bunches is quite small because of
destructive interference, for example, $[1-J_{1}(p_{1})]^{2}=0.175$ if
$p_{1}=1.8$; $[1-J_{2}(p_{2})]^{2}=0.26$ if $p_{2}=3.1$; and $[1-J_{3}%
(p_{3})]^{2}=0.32$ if $p_{3}=4.2$, i.e., almost an order of magnitude smaller
with respect to the pulse maxima. Qualitatively the appearance of bunches can
be understood from  the time evolution analysis of the phase difference of
the scattered field and the comb, which is $\psi_{m}(t)+\pi$ if $S_{m}(t)<0$.
The phase $\psi_{m}(t)$ evolves almost linearly as $(m+p)\Omega t+C$ during
the pulse (pulses) formation around $\Omega t_{p} = (2n+1)\pi$ ($C$ is constant
within each time interval), and the phase evolution almost stops around
$\Omega t_{s} = 2n\pi$, see SM for details.

It is interesting to notice that tuning in resonance the $m=-1$ component (if
$\varphi=0$) shifts the position of the pulses with respect to the case of
$m=+1$ such that maxima become minima and vice versa. Such a difference is
explained by the fact that the amplitude of the $m=-1$ component of the comb
is proportional to $J_{-1}(p)=-J_{1}(p)$ and hence the amplitude of the
antiphase scattered field [proportional to $S_{-1}(t)$] becomes positive. The
details how to move the radiation field from one time-bin to the other time-bin
by changing the phase $\varphi$ and how this effect can be used to create time-bin
qubits and qudits are discussed in SM.

\textbf{Experiment.} The details of the experimental setup are described in
Refs. \cite{Shakhmuratov09,Shakhmuratov11}. Recently we developed a new scheme
of photon counts selection and reported our first observation of photon
shaping into a series of short single-pulse bunches \cite{Vagizov}. This
transformation is performed by tuning the radiation source in resonance with
the first sideband $m=1$. Below we sketch out the experimental scheme.

The radiation source is a radioactive $^{57}$Co in rhodium matrix. The
absorber is a 25-$\mu$m-thick stainless-steel foil with a natural abundance
($\sim$ 2\%) of $^{57}$Fe. Optical depth of the absorber is $T_{a}=5.18$. The
stainless-steel foil is glued on the polyvinylidene fluoride piezo-transducer
that transforms the sinusoidal signal from radio-frequency generator into the
uniform vibration of the foil. The tuning of the source in resonance for the
preselected component of the spectrum is performed by M\"{o}ssbauer transducer
working/running in constant velocity mode. The source is attached to the
holder of the M\"{o}ssbauer transducer causing Doppler shift of the radiation
field. Two detectors, D1 and D2, are used in data acquisition scheme. D1
(shielded by copper foil) detects only heralding 122 keV photons in a cascade
decay, 122 keV and 14.4 keV, of $^{57}$Co. This detector starts the clock.
Detection of 14.4 keV photon by D2 stops the clock. In this time-delayed
coincidence count technique we reconstruct the time evolution of the photon
wave packet transmitted through the resonant absorber. Since time $t_{0}$ of
the formation of 14.4 keV state nucleus in the source is random, we select
only those counts of the heralding 122 keV photons, which are detected within
short time interval $\Delta t$ around time $t_{ph}$ satisfying the relation
$\Omega t_{ph}=\varphi+2\pi n$, where $n$ is integer. This selection secures
that the phase of the absorber vibration is always the same for all detected
photons. Since small time window of count selection $\Delta t$ is not zero, we
have to average the theoretical expressions for the signal $P(t)$ over small
jitter $\Delta\varphi$ of phase $\varphi$ caused by finite value of $\Delta
t$. \begin{figure}[ptb]
\resizebox{0.35\textwidth}{!}{\includegraphics{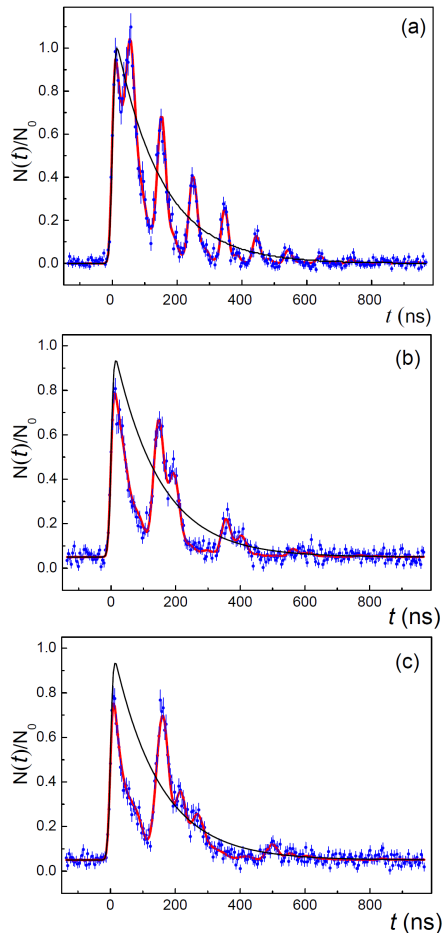}}\caption{(color on
line) Time dependence of the photon counts $N(t)$ at the exit of the absorber
vibrating with the frequency $\Omega=10.2$ MHz (a), $\Omega=4.79$ MHz (b), and
$\Omega=2.94$ MHz (c). $N(t)$ is normalized to the value $N_{0}$ without resonant
absorbtion if the contribution of nonresonant photons with recoil is subtracted.
The frequency of the radiation field $\omega_{s}$ is
tuned in resonance with the first sideband $\omega_{a}+\Omega$ (a), the second
sideband $\omega_{a}+2\Omega$ (b), and the third sideband $\omega_{a}+3\Omega$
(c). Solid line (in red) corresponds to the theoretical fitting, which takes
into account the phase jitter. The modulation index, the phase, and the phase
jitter $\Delta\varphi$ are $p=1.8$, $\varphi=0$, and $\Delta\varphi=\pi/2$
(a), $p=3.08$, $\varphi=0$, and $\Delta\varphi=\pi/3$ (b), $p=4.21$,
$\varphi=-\pi/10$, and $\Delta\varphi=\pi/5$ (c). The dots with error bars (in
blue) show experimental points. Thin line (in black) shows the probability
time dependence without absorber.}%
\label{fig:2}%
\end{figure}\begin{figure}[ptbptb]
\resizebox{0.35\textwidth}{!}{\includegraphics{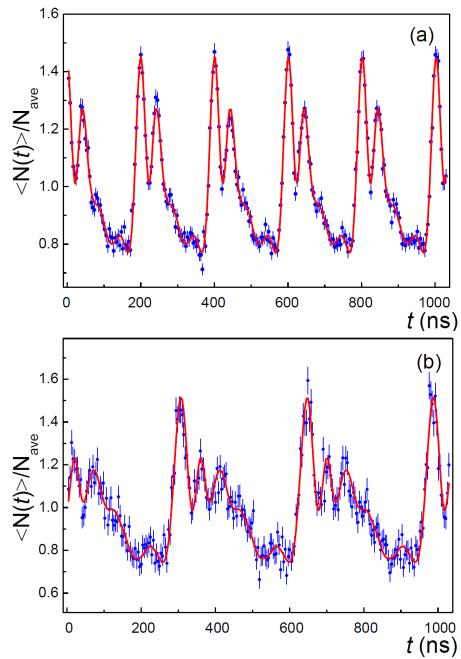}}\caption{Time
dependence of the $t_{0}$ averaged detection probability of a photon,
$\langle N(t)\rangle\sim\langle P(t)\rangle_{t_{0}}$, at the exit of the
vibrating absorber. $N_{ave}$ is the averaged count rate. The
experimental conditions and notations in plots (a) and (b) are the same
as in plots (b) and (c) in Fig. 2, respectively. The experimental results
are fitted to the exact calculation of the integral in Eq. (\ref{Eq7}),
see the details in SM.}%
\label{fig:3}%
\end{figure}

Time resolution of the electronics in our setup is 8 ns and hence time
structures shorter than 8 ns would be not resolved. In Ref. \cite{Vagizov} we
modulated the absorber with frequency $\Omega=10.2$ MHz and tuned the
radiation field in resonance with the first satellite. We observed pulses as
short as 30 ns, which are artificially broadened due to $\Delta\varphi\neq 0$.
If we tune the source in resonance with the second or third
satellite and keep the same frequency of modulation $\Omega=10.2$ MHz, the
pulses will be 2 or 3 times shorter, respectively (see Fig. 1), which makes
difficult to resolve pulses within the bunches. To be able to resolve the
content of the pulse bunches, for example, consisting of two or three pulses,
we had to reduce $\Omega$ two or three times, respectively, compared to the
modulation frequency, used in our first experiment. The experimental results
of the detecting of pulse bunching are presented in Fig. 2, where time
dependence of the number of counts $N(t)$, normalized to the maximum value
$N_{0}$ without absorber, is shown. The ratio $N(t)/N_{0}$ is proportional to
the probability $P(t)$. The details of fitting procedure are described in Ref.
\cite{Vagizov}.

To avoid smearing out of the pulses within the bunch we performed experiments
where time $t_{0}$ is still random but we count time delay of 14.4 keV photon
detection by the detector D2 with respect to the fixed moments of
time when the modulation phase $\varphi$ is the same (differing only in $2\pi
n$). Thus, we don't use detector D1 and time delay of the detection of 14.4
keV photon, $t_{stop}$, is counted with respect to a fixed $t_{start}$, which
is not random. In this way we escape an artificial phase jitter $\Delta\varphi$
inherent to the first scheme of the experiment. What we measure in the
modified scheme of experiment is the probability $P(t)$, integrated over time
$t_{0}$, which varies from $-\infty$ to $t$, i.e.,
\begin{equation}
\left\langle P(t-t_{0})\right\rangle _{t_{0}}=2\gamma_{s}%
%TCIMACRO{\dint \limits_{-\infty}^{t}}%
%BeginExpansion
{\displaystyle\int\limits_{-\infty}^{t}}
%EndExpansion
P(t-t_{0})dt_{0}. \label{Eq7}%
\end{equation}
Calculation of this integral for the analytical approximation, Eq.
(\ref{Eq5}), gives
\begin{equation}
\left\langle P(t)\right\rangle _{t_{0}}=1-2V_{m}(p)\cos\psi_{m}(t)+U_{m}(p),
\label{Eq8}%
\end{equation}
where $V_{m}(p)=J_{m}(p)[1-\exp(-T_{a}/4)]$, $U_{m}(p)=V_{m}^{2}(p)+J_{m}%
^{2}(p)\exp(-T_{a}/2)[I_{0}(T_{a}/2)-1]$, and $I_{0}(T_{a}/2)$ is the modified
Bessel function of zero order. We notice that for a single line absorber (not
vibrating) a steady state transmission of resonant gamma-quanta is
proportional to the function $\exp(-T_{a}/2)I_{0}(T_{a}/2)$, where $T_{a}$ is
the optical depth of the absorber. Therefore, we conclude that part of the
term $U_{m}(p)$, which is proportional to this function, describes in Eq.
(\ref{Eq8}) a steady state transmission of the resonant $m$-component of the
frequency comb, within the assumption that other components pass through
without any change. The exact expression for $\left\langle P(t-t_{0}%
)\right\rangle _{t_{0}}$, obtained without analytical approximation
(\ref{Eq5}), looks very complicated and does not allow simple interpretation
and analytical analysis, see Ref. \cite{Ikonen}. Meanwhile, our approximation
slightly deviates from this exact result. However, if the contribution of two
neighboring sidebands, blue and red detuned from resonant component, are taken
into account, then this difference becomes negligible (see SM).

Figure 3 demonstrates the results of our time-delayed measurements with
respect to a fixed phase of the vibrations. In spite of poor time resolution
of our electronics the pulses within bunches are clearly seen.

Concluding, we demonstrate a method how to shape single-photon wave packets
into bunches of pulses by transmission through the vibrated single-line
absorber. We control the number of pulses in a bunch and their timing by
tuning the absorber to the proper radiation sideband induced by the absorber
vibration.

This work was partially funded by the Russian Foundation for Basic Research
(Grant No. 12-02-00263-a), the Program of Competitive Growth of Kazan Federal
University funded by the Russian Government, the RAS Presidium Program
"Quantum Mesoscopic and Disordered Systems", the National Science
Foundation (Grant No. PHY-1307346), and the Robert A. Welch Foundation
(Award A-1261). We are grateful to Alexey Kalachev for the fruitful
discussions.

\end{document}

\title{Supplemental Material for:\\Transformation of a single photon field into bunches of pulses}
\author{R. N. Shakhmuratov}
\affiliation{Kazan Physical-Technical Institute, Russian Academy of Sciences,
10/7 Sibirsky Trakt, Kazan 420029 Russia}
\affiliation{Kazan Federal University, 18 Kremlyovskaya Street, Kazan 420008 Russia}
\author{F.G. Vagizov}
\affiliation{Kazan Physical-Technical Institute, Russian Academy of Sciences,
10/7 Sibirsky Trakt, Kazan 420029 Russia}
\affiliation{Kazan Federal University, 18 Kremlyovskaya Street, Kazan 420008 Russia}
\affiliation{Department of Physics and Institute for Quantum Studies, TAMU, College
Station, Texas 77843-4242, USA}
\author{V.A. Antonov}
\affiliation{Kazan Federal University, 18 Kremlyovskaya Street, Kazan 420008 Russia}
\affiliation{Institute of Applied Physics of the Russian Academy of Sciences, Nizhny
Novgorod 603950, Russia}
\affiliation{N.I. Lobachevsky State University of Nizhny Novgorod,
Nizhny Novgorod, 603950, Russia}
\author{Y.V. Radeonychev}
\affiliation{Institute of Applied Physics of the Russian Academy of Sciences, Nizhny
Novgorod 603950, Russia}
\affiliation{N.I. Lobachevsky State University of Nizhny Novgorod,
Nizhny Novgorod, 603950, Russia}
\author{O. Kocharovskaya}
\affiliation{Department of Physics and Institute for Quantum Studies, TAMU, College
Station, Texas 77843-4242, USA}
\date{{ \today}}

\begin{abstract}
We show here that taking into account the contribution of the nearest
satellites of the resonant component removes misfit of our analytical
approximation with the exact result for the probability amplitude of the
photon, transmitted through the vibrating absorber. We analyze time evolution
of the phase difference of the scattered field and the comb. We discuss the
scheme how single and two-pulse bunches can be used to simulate spin $1/2$ qubit and
ququad.

\end{abstract}
\maketitle

\section{Contribution of nonresonant components}

If the modulation frequency $\Omega$ is much larger than the halfwidth of the
absorption line $\gamma_{a}$ and only one spectral component of the frequency
comb is tuned in resonance with the absorber then one can neglect the
interaction with other nonresonant components. Such an idealization works quite
well if the spectrum of a single-line radiation source has rapidly falling
tails as it is inherent, for example, to the Gaussian spectrum. In case of
heralded single photons the spectrum of the radiation field is $a(\omega
)=i/(\omega-\omega_{s}+i\gamma_{s})$, see, for example Ref.
\cite{Shakhmuratov09}. This spectrum has long tails falling as $\sim
1/(\omega-\omega_{s})$. They appear because the front of the photon wave packet
has sharply rising leading edge. Therefore many satellites start wringing immediately
after this front comes, however with small amplitudes. As a result the
approximate Eq. (5) describes the modulation of a single photon field with
small misfit compared with exact Eq. (6) in the main text of the paper. Below
we improve fitting by taking into account the interaction with the nearest
satellites of the resonant component.

\subsection{Nonaveraged probability amplitude}

The propagation of a single line radiation field with carrier frequency
$\omega_{s}$ through a single line absorber with resonant frequency
$\omega_{a}$ is described by Eq. (6) of the main part of the paper. It is
derived by the convolution of the incident field amplitude with the response
function (Green function) of a single line absorber, which is
\begin{equation}
R(t)=\delta(t)-\Theta(t)e^{-(i\omega_{a}+\gamma_{a})t}bj_{1}\left(  bt\right)
, \label{Eq1}%
\end{equation}
where $\delta(t)$ is the Dirac delta function (see, for example, Refs.
\cite{Shakhmuratov09,Ikonen}). According to Eq. (6) of the main part of the
manuscript the coherently scattered field amplitude is%
\begin{equation}
a_{sct}(t-t_{0},l)=-\Theta(t-t_{0})b%
%TCIMACRO{\dint \limits_{0}^{t-t_{0}}}%
%BeginExpansion
{\displaystyle\int\limits_{0}^{t-t_{0}}}
%EndExpansion
a_{A}(t-t_{0}-\tau,l)j_{1}(b\tau)e^{-\gamma_{a}\tau-i\omega_{a}\tau}d\tau.
\label{Eq2}%
\end{equation}
If the $m$-component of the frequency comb, Eq. (1) of the main part of the
manuscript, is in resonance with the absorber, i.e., $\omega_{s}=\omega
_{a}+m\Omega$ and $\gamma_{s}=\gamma_{a}$, then the amplitude of the
coherently scattered field for this component is reduced to%
\begin{equation}
a_{m}(t-t_{0},l)=a_{L}(t-t_{0},l)S_{m}(t-t_{0})e^{im(\Omega t+\varphi)},
\label{Eq3}%
\end{equation}
where for simplification of the notations we drop index $sct$. The
contribution of the satellites $\omega_{s}-(m\pm n)\Omega$, where
$n=1,2,3,...$, was not taken into account in Eq. (5) of the the main part of
the manuscript. According to Eq. (\ref{Eq2}) the amplitudes of the fields,
scattered by the satellites are%
\begin{equation}
a_{m\pm n}(t-t_{0},l)=-a_{L}(t-t_{0},l)J_{m\pm n}(p)e^{i(m\pm n)(\Omega
t+\varphi)}b%
%TCIMACRO{\dint \limits_{0}^{t-t_{0}}}%
%BeginExpansion
{\displaystyle\int\limits_{0}^{t-t_{0}}}
%EndExpansion
j_{1}(b\tau)e^{-(\gamma_{a}-\gamma_{s})\tau\mp in\Omega\tau}d\tau. \label{Eq4}%
\end{equation}
Then, the exact result, Eq. (6) of the the main part of the manuscript, can be
expressed as follows%
\begin{equation}
a_{ext}(t,l)=a_{A}(t,l)+\sum_{n=-\infty}^{+\infty}a_{m+n}(t,l), \label{Eq5}%
\end{equation}
where $t_{0}=0$. Comparison of the exact result with the approximation, when
only two nearest sattelites ($n=\pm1$) of the resonant component ($n=0$) are
taken into account, is shown in Fig. 1. In this case misfit is almost
negligible. \begin{figure}[ptb]
\resizebox{0.4\textwidth}{!}{\includegraphics{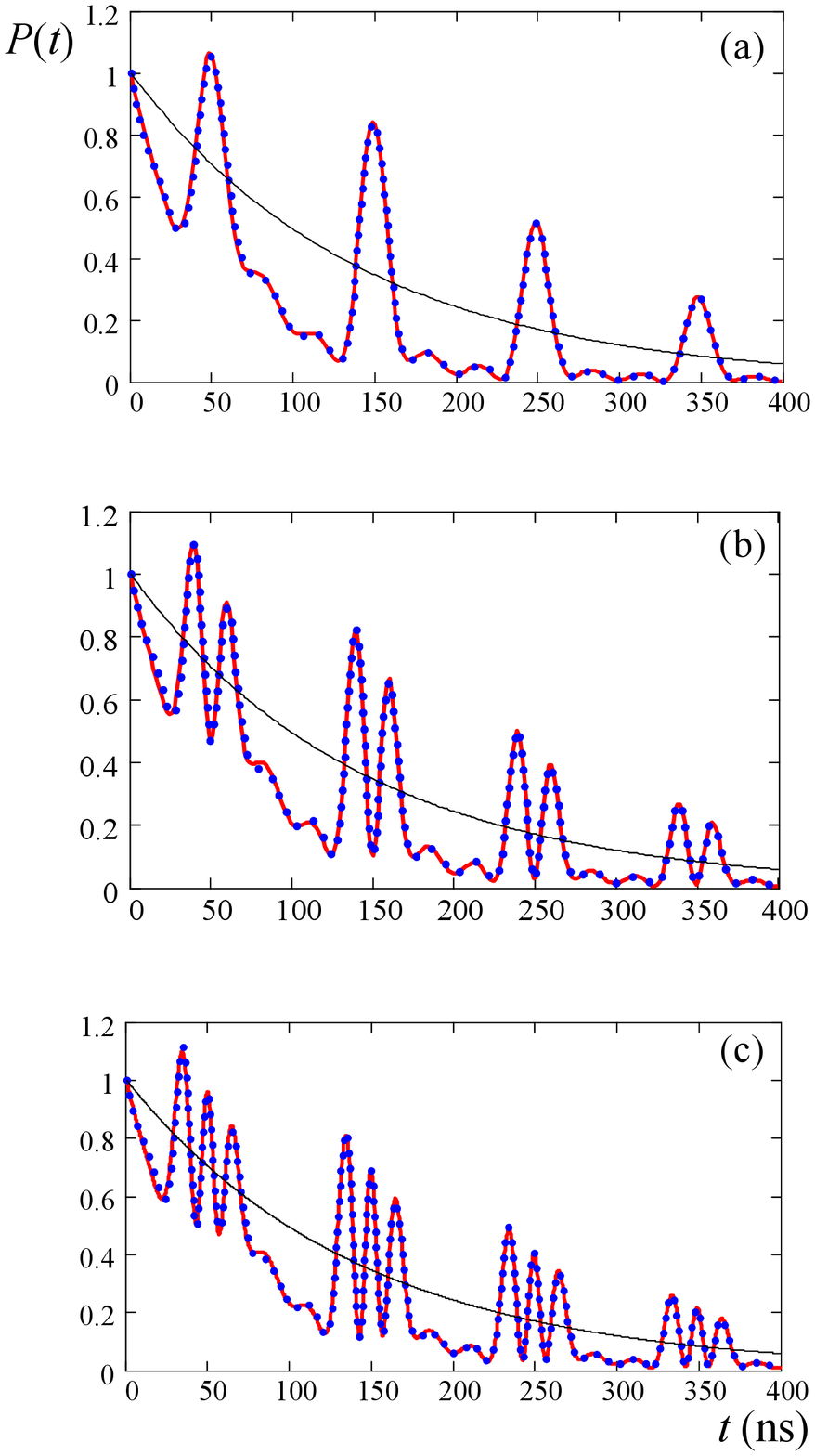}}\caption{(color on
line) Comparison of the exact result for the probability $P(t)$, derived from
Eq. (6) in the main part of the manuscript (solid line in red), with that,
which is obtained from Eq. (\ref{Eq5}), where only the resonant component,
$n=0$, and two nearest satellites, $n=\pm1$, are taken into account (dotted
line in blue). The parameters and notations are the same as in Fig.1 (a-c) of
the main part of the manuscript.}%
\label{fig:1}%
\end{figure}\begin{figure}[ptbptb]
\resizebox{0.4\textwidth}{!}{\includegraphics{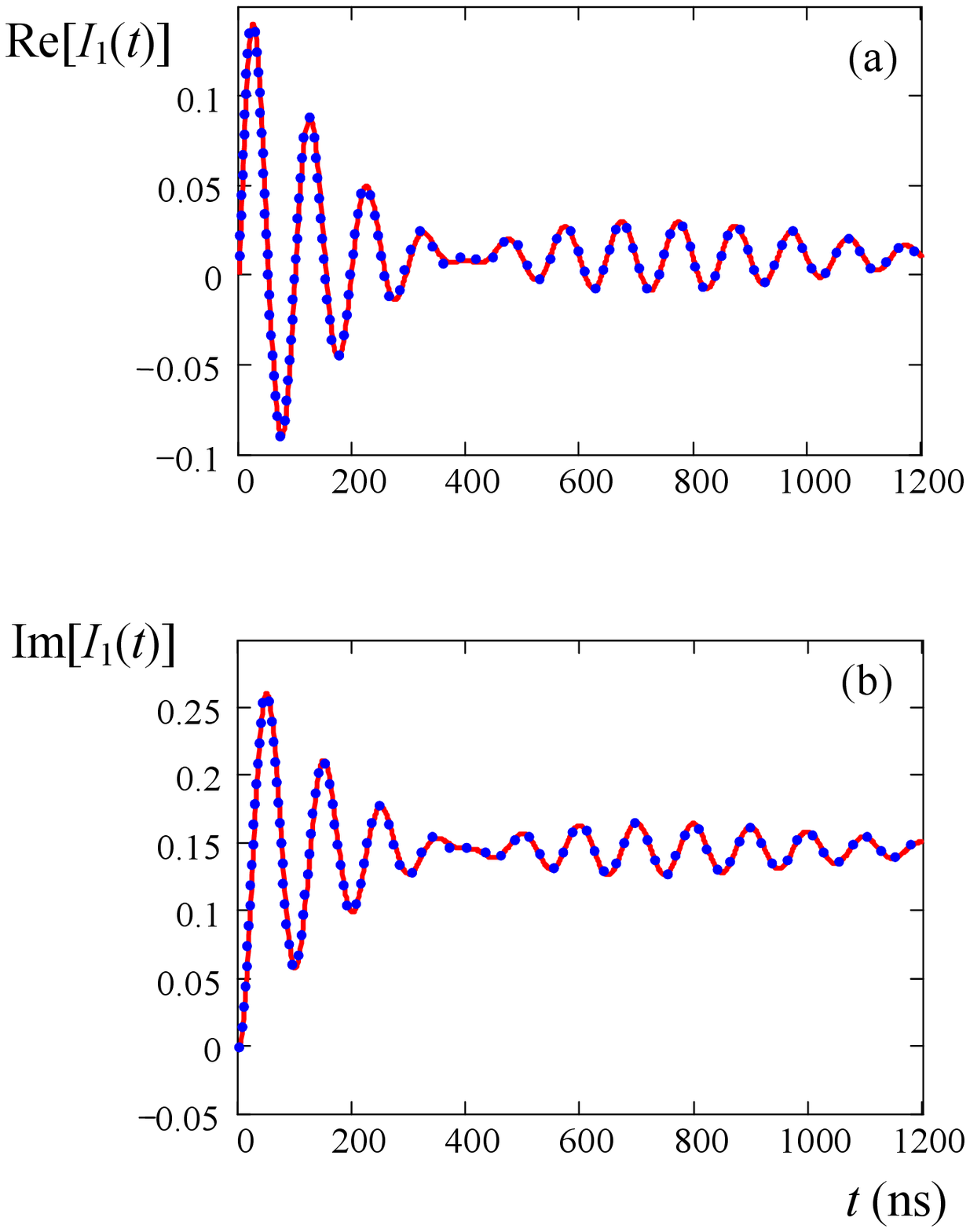}}\caption{(color on
line) Comparison of the time dependencies of the real (a) and imaginary (b)
parts of the integral $K_{1}(t)$, Eq. (\ref{Eq6}), (solid line in red) with
the equation (\ref{Eq7}), where only the first term in the sum, Eq.
(\ref{Eq8}), is taken into account (dotted line in blue). The parameters are
$b=1.47$ MHz and $\Omega=10$ MHz.}%
\label{fig:2}%
\end{figure}To estimate the contribution of the satellites we calculated the
integral%
\begin{equation}
K_{\pm n}(t)=b%
%TCIMACRO{\dint \limits_{0}^{t}}%
%BeginExpansion
{\displaystyle\int\limits_{0}^{t}}
%EndExpansion
j_{1}(b\tau)e^{\pm in\Omega\tau-(\gamma_{a}-\gamma_{s})\tau}d\tau\label{Eq6}%
\end{equation}
in Eq. (\ref{Eq4}) with the help of the method, presented in Refs.
\cite{Lynch,Crisp,Varoquaux,Kuznetsova}. The result is
\begin{equation}
K_{\pm n}(t)=1-e^{-ib/[\pm n\Omega+i(\gamma_{a}-\gamma_{s})]}+M_{\pm n}(t),
\label{Eq7}%
\end{equation}%
\begin{equation}
M_{\pm n}(t)=e^{\pm in\Omega t-(\gamma_{a}-\gamma_{s})t}\sum_{k=1}^{\infty
}\left[  \frac{-ib}{\pm n\Omega+i(\gamma_{a}-\gamma_{s})}\right]  ^{k}%
j_{k}(bt), \label{Eq8}%
\end{equation}
where $j_{k}(bt)=J_{k}\left(  2\sqrt{bt}\right)  /(bt)^{k/2}$ and $J_{k}(x)$
is the Bessel function of the $k$-th order. If $n\Omega\gg b$, then the
smallness of the satellites contribution is of the order of $b/n\Omega$. For
example, when $b/\Omega=0.146$, $n=1$, and $\gamma_{a}\approx\gamma_{b}$, then
it is already fine approximation if one takes into account only the first term
in the sum $M_{\pm 1}(t)$ in Eq. (\ref{Eq8}), which is proportional to
$b/\Omega$, see Fig. 2.

\subsection{Averaged probability amplitude}

The averaged probability amplitude $\left\langle P(t-t_{0})\right\rangle
_{t_{0}}=\left\langle N(t)\right\rangle $, Eq. (\ref{Eq7}) in the main part of
the paper, is described by the equation (see Refs. \cite{Ikonen,Kuznetsova})%
\begin{multline}
\left\langle N(t)\right\rangle =\operatorname{Re}\Bigg [1-2f_{s}bF_{+}%
(t)\int_{-\infty}^{t}dt^{\prime}j_{1}[b(t-t^{\prime})]/F_{+}(t^{\prime})\\
+2f_{s}b^{2}e^{-2\gamma_{a}t}\int_{-\infty}^{t}dt^{\prime}F_{-}(t^{\prime
})j_{1}[b(t-t^{\prime})]\int_{-\infty}^{t^{\prime}}dt^{\prime\prime}%
j_{1}[b(t-t^{\prime\prime})]/F_{+}(t^{\prime\prime})\Bigg ]. \label{Eq9}%
\end{multline}
where $f_{s}$ is the recoilless fraction of the source photons and%
\begin{equation}
F_{\pm}(t)=\exp\left[  -(\gamma_{s}\pm\gamma_{a})t-i(\omega_{a}-\omega
_{s})t-ip\sin(\Omega t)\right]  . \label{Eq10}%
\end{equation}
If $\gamma_{s}=\gamma_{a}=\gamma$, then Eq. (\ref{Eq9}) can be simplified as
follows%
\begin{multline}
\left\langle N(t)\right\rangle =1-2f_{s}b\int_{0}^{\infty}dt^{\prime}%
j_{1}(bt^{\prime})e^{-2\gamma t^{\prime}}\cos\left[  \phi(t)-\phi(t-t^{\prime
})-\Delta\omega t^{\prime}\right] \\
+2f_{s}b^{2}\int_{0}^{\infty}dt^{\prime}\int_{0}^{t^{\prime}}dt^{\prime\prime
}j_{1}(bt^{\prime})j_{1}(bt^{\prime\prime})e^{-2\gamma t^{\prime}}\cos\left[
\phi(t-t^{\prime})-\phi(t-t^{\prime\prime})+\Delta\omega(t^{\prime}%
-t^{\prime\prime})\right]  . \label{Eq11}%
\end{multline}
where $\Delta\omega=\omega_{s}-\omega_{a}$ is the resonant detuning and
$\phi(t)=p\sin(\Omega t+\varphi)$ is the phase modulation of the radiation
field in the reference frame of the vibrating absorber.

Equations (\ref{Eq9}) and (\ref{Eq11}) are hard to analyze analytically. On
the contrary, analytical approximation, given in Eq. (\ref{Eq8}) of the main
part of the paper, helps to estimate the periodicity of pulses, the values of
their maxima $\left\langle N_{m}\right\rangle _{\max}$, and the intensity
level of the dark windows, $\left\langle N_{m}\right\rangle _{\min}$, where
index $m$ indicates that $m$-th component is in resonance. According to the
analytical approximation these values are%
\begin{equation}
\left\langle N_{m}\right\rangle _{\max}=\left[  1+V_{m}(p)\right]  ^{2}%
+J_{m}^{2}(p)(\left\langle N\right\rangle _{res}-e^{-T_{a}/2}), \label{Eq12}%
\end{equation}%
\begin{equation}
\left\langle N_{m}\right\rangle _{\min}=\left[  1-V_{m}(p)\right]  ^{2}%
+J_{m}^{2}(p)(\left\langle N\right\rangle _{res}-e^{-T_{a}/2}), \label{Eq13}%
\end{equation}
where $\left\langle N\right\rangle _{res}=\exp(-T_{a}/2)I_{0}(T_{a}/2)$ is
proportional to the number of counts per unit time at the output of the
single-line absorber if it is tuned in resonance with the source. Comparison
of the analytical approximation, Eq. (8) in the main part of the paper, with
the exact expression, Eq. (\ref{Eq11}), for the number of counts at the output
of the vibrating absorber, is shown in Fig. 3, along with the values
$\left\langle N_{m}\right\rangle _{\max}$, $\left\langle N_{m}\right\rangle
_{\min}$, and $\left\langle N\right\rangle _{res}$, which are $\left\langle
N_{1}\right\rangle _{\max}=2.089$, $\left\langle N_{1}\right\rangle _{\min
}=0.397$ for $\Delta=\Omega$ and $p=1.8$; $\left\langle N_{2}\right\rangle
_{\max}=1.877$, $\left\langle N_{2}\right\rangle _{\min}=0.463$ for
$\Delta=2\Omega$ and $p=3.1$; $\left\langle N_{3}\right\rangle _{\max}=1.768$,
$\left\langle N_{3}\right\rangle _{\min}=0.504$ for $\Delta=3\Omega$ and
$p=4.2$. These values are obtained if modulation frequency is $\Omega=10$ MHz
and $T_{a}=5.2$, when $\left\langle N\right\rangle _{res}=0.264$. The maximum
intensity of the pulses exceeds almost two times the radiation intensity
without absorber. Minimum intensity in the dark windows $\left\langle
N_{m}\right\rangle _{\min}$ is almost an order of magnitude smaller than the
maximum pulse intensity $\left\langle N_{m}\right\rangle _{\max}$ and slightly
exceeds the intensity level of the radiation field, transmitted through the
resonant absorber not vibrating, $\left\langle N\right\rangle _{res}$. However
their difference $\left\langle N_{m}\right\rangle _{min}-\left\langle
N\right\rangle _{res}$ rises with increase of the value of the modulation
index from $p=1.8$ to $p=4.2$. \begin{figure}[ptb]
\resizebox{0.5\textwidth}{!}{\includegraphics{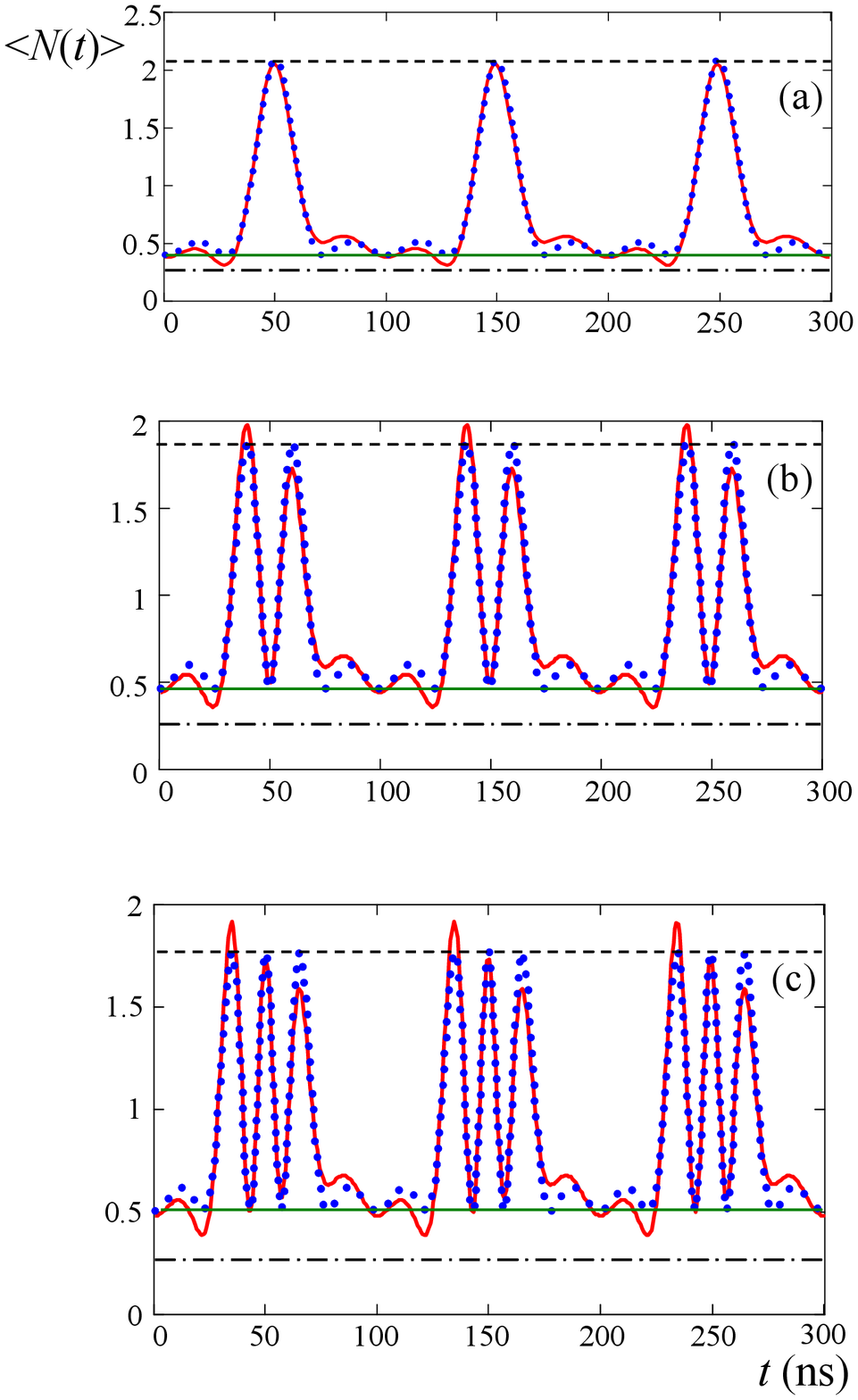}}\caption{(color on
line) Time dependence of the photon counts, averaged over $t_{0}$, for
$\Delta=\Omega$ and $p=1.8$ (a), $\Delta=2\Omega$ and $p=3.1$ (b),
$\Delta=3\Omega$ and $p=4.2$ (c). Solid line (in red) represents the exact
result and dotted line (in blue) - analytical approximation, Eq. (8) from the
main part of the paper. Dashed black line shows the level $\langle
N_{m}\rangle_{max}$, solid line (in green) - $\langle N_{m}\rangle_{min}$, and
dash-dotted line represents $\langle N\rangle_{res}$. Other parameters are
defined in the text.}%
\label{fig:3}%
\end{figure}\begin{figure}[ptbptb]
\resizebox{0.4\textwidth}{!}{\includegraphics{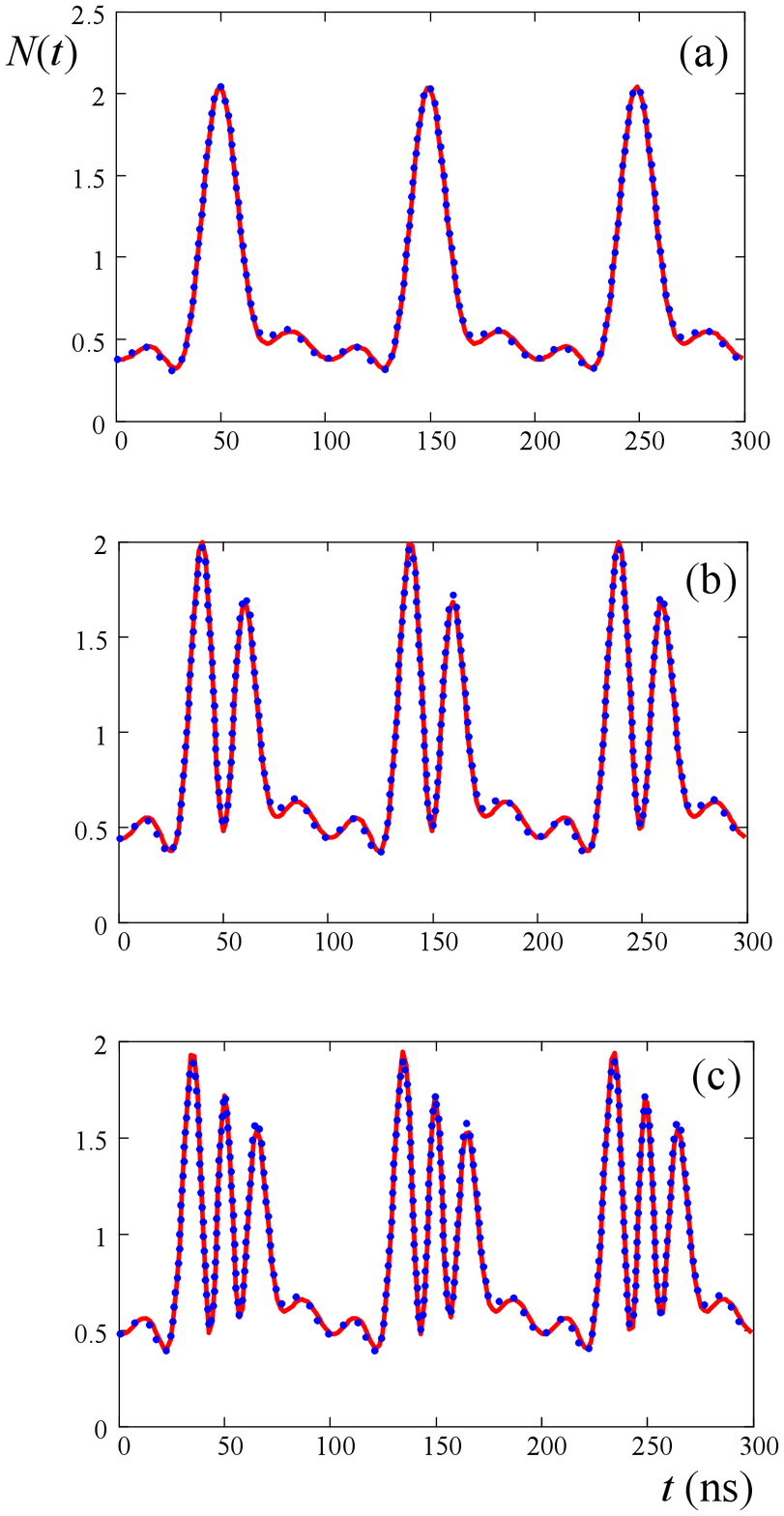}} \caption{(color on
line) Comparison of the exact result (solid line in red) for time dependence
of the photon counts, averaged over $t_{0}$, for $\Delta=\Omega$ and $p=1.8$
(a), $\Delta=2\Omega$ and $p=3.1$ (b), $\Delta=3\Omega$ and $p=4.2$ (c) with
analytical approximation Eq. (\ref{Eq15}) (dotted line in blue). Other parameters
are the same as in Fig.1 (a-c) of
the main part of the manuscript.}%
\label{fig:4}%
\end{figure}
\begin{figure}[ptb]
\resizebox{0.4\textwidth}{!}{\includegraphics{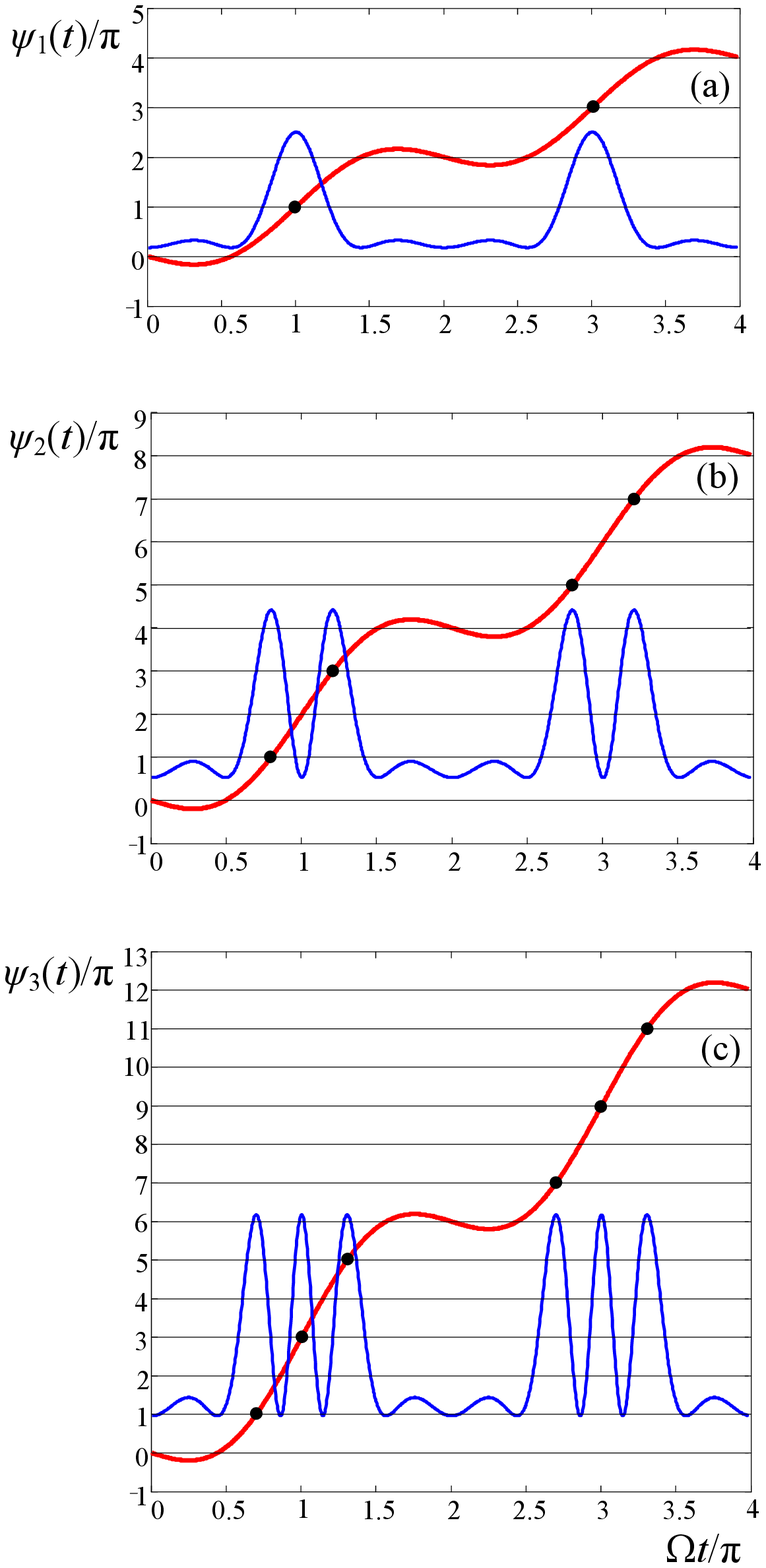}}\caption{(color on
line) Time evolution of the phase difference of the comb and resonantly
scattered field component, $\psi_{m}(t)$, for $m=1$ (a), $m=2$ (b), and $m=3$
(c), thick line in red. Black circles indicate the points when $\psi
_{m}(t)=(2n+1)\pi$. The values of the modulation index are taken the same as
in Fig. 1. The modulation phase is $\varphi=0$. Thin solid line in blue shows
the formation of pulses according to Eq. (\ref{Eq5}) in the main part of the
manuscript. For visualization the exponential factor is removed and time
dependent Bessel function is set equal to zero. The amplitudes of the pulses
are scaled to fit a half of each plate.}%
\label{fig:5}%
\end{figure}

It is possible to improve the analytical approximation, given in Eq. (8) of
the main part of the paper, if the contribution of two nearest satellites of
the resonant component are taken into account, i.e.,%
\begin{equation}
a_{apx}(t,l)=a_{A}(t,l)+a_{m}(t,l)+a_{m+1}(t,l)+a_{m-1}(t,l). \label{Eq14}%
\end{equation}
In the corresponding probability amplitude $P_{apx}(t)=\left\vert
a_{apx}(t,l)\right\vert ^{2}$ we can make further simplification neglecting
the terms $\left\vert a_{m+1}(t,l)+a_{m-1}(t,l)\right\vert ^{2}$ and
$2\operatorname{Re}\left\{  \left[  a_{m+1}(t,l)+a_{m-1}(t,l)\right]
a_{L}(t,l)J_{m}\left(  2\sqrt{bt}\right)  e^{im(\Omega t+\varphi)}\right\}  $
since their contribution into the averaged probability $\left\langle
P_{apx}(t-t_{0})\right\rangle _{t_{0}}$ is small. The contribution of other
terms results in expression%
\begin{equation}
\left\langle N_{apx}(t)\right\rangle =\left\langle N_{m}(t)\right\rangle
-\left\langle C_{m+1}(t)\right\rangle -\left\langle C_{m-1}(t)\right\rangle ,
\label{Eq15}%
\end{equation}
where $\left\langle N_{m}(t)\right\rangle =\left\langle P(t-t_{0}%
)\right\rangle _{t_{0}}$ is defined in Eq. (\ref{Eq8}) of the main part of the
paper and
\begin{multline}
\left\langle C_{m\pm1}(t)\right\rangle =2J_{m\pm1}(p)\Bigg \{\cos\psi_{m\pm
1}(t)-e^{-B}\cos\left[  \psi_{m\pm1}(t)\pm D\right] \\
-J_{m}(p)\left[  \cos(\Omega t+\varphi)-e^{-B}\cos(\Omega t+\varphi+D)\right]
\Bigg \}, \label{Eq16}%
\end{multline}%
\begin{equation}
B+iD=\frac{b}{2\gamma-i\Omega}. \label{Eq17}%
\end{equation}
It is easy to show that the contribution of terms $\left\langle C_{m\pm
1}(t)\right\rangle $ is as small as $2b\gamma/\Omega^{2}$ if $\Omega\gg
2\gamma$ and $\Omega\gg b$. However, in spite of the smallness of the
corrections, Eq. (\ref{Eq15}) describes much better the formation of pulses
than the analytical approximation Eq. (8) in the main part of the manuscript,
see Fig. 4, where Eq. (\ref{Eq15}) is compared with the exact result. Misfit
between them is almost negligible.

\section{Phase evolution and formation of pulses and dark windows}

Time dependence of phase $\psi_{m}(t)$ is shown in Fig. 5 for $m=1$ (a),
$m=2$ (b), and $m=3$ (c) if $\varphi=0$ and modulation index $p$ has optimal
value for each $m$. The phase $\psi_{m}(t)$ evolves almost linearly as
$(m+p)\Omega t+C$ during the pulse (pulses) formation around
$\Omega t_{p}=(2n+1)\pi$ ($C$ is constant within each time interval), and the
phase evolution almost stops around $\Omega t_{s}=2n\pi$. Durations of linear
time evolution and the phase stopping intervals are nearly equal each other
and they are nearly confined within the time intervals
$(\Omega t_{p}-\pi/2,\Omega t_{p}+\pi/2)$ and $(\Omega t_{s}-\pi/2,\Omega t_{s}+\pi/2)$,
respectively. The phase stops could be explained by "destructive interference"
of two terms in expression for the phase $\psi_{m}(t)$ at the optimal values of
the modulation indexes since at the stops the phase evolution is approximated
as $(m-p)\Omega t+C$. Actually these "phase stopping" periods are the periods
when time evolution of $\psi_{m}(t)$ changes the slope from $(m+p)\Omega t$
to $(m-p)\Omega t$. Since for the optimal values of the modulation index $p$
we have $p>m$ and approximately the relation $p\approx m+1$ is valid, then the
slope of the phase change during formation of pulses is $(2m+1)\Omega t$ and
this slope is negative, $-\Omega t$, during the dark windows.

\section{Operations with time bin qubits}

Here we consider two algorithms how pulse bunching can be used to create and
operate with time bin qubits. Assume that the phase of absorber vibrations is
zero, $\varphi=0$, and we tune the radiation source in resonance with the
first satellite, $\omega_{s}=\omega_{a}+\Omega$. Then the pulses are formed at
the moments of time $t_{p}=(2n+1)T_{v}/2$, where $T_{v}$ is the period of the
vibrations and $n=0,1,2...$ The dark windows are formed around the moments of
time $t_{d}=nT_{v}$. \ The illustration of these pulses and dark windows is
given in Fig. 6a. In the bottom panel, Fig. 6d, the evolution of the radiation
phase, $\sin(\Omega t)$, is shown. It is divided into bins A and B. The A bins
are centered at times $t_{p}$ where the pulses are formed due to constructive
interference of the incident and coherently scattered radiation fields. The B
bins are centered at times $t_{d}$ where dark windows appear due to
destructive interference of the incident and coherently scattered radiation
fields. The length of these bins is equal to half a period of the vibrations,
$T_{v}/2$. Below we assume that we have a local oscillator, for example, a
generator producing the voltage oscillating according to the function
$\sin(\Omega t+\varphi_{lo})$ with phase $\varphi_{lo}=0$. The same oscillator
generates mechanical vibrations of the absorber with tunable phase $\varphi$.
If $\varphi=0$, all the pulses are formed at the output of the absorber in A
bins and dark window are located in B bins (see Fig. 6a). If $\varphi=\pi$ the
pulses are generated in B bins, while dark windows are located in A bins (see
Fig. 6b). If $\varphi=\pi/2$, then the radiation field is equally distributed
among A and B bins (see Fig. 6c).
\begin{figure}[ptbptbptb]
\resizebox{0.4\textwidth}{!}{\includegraphics{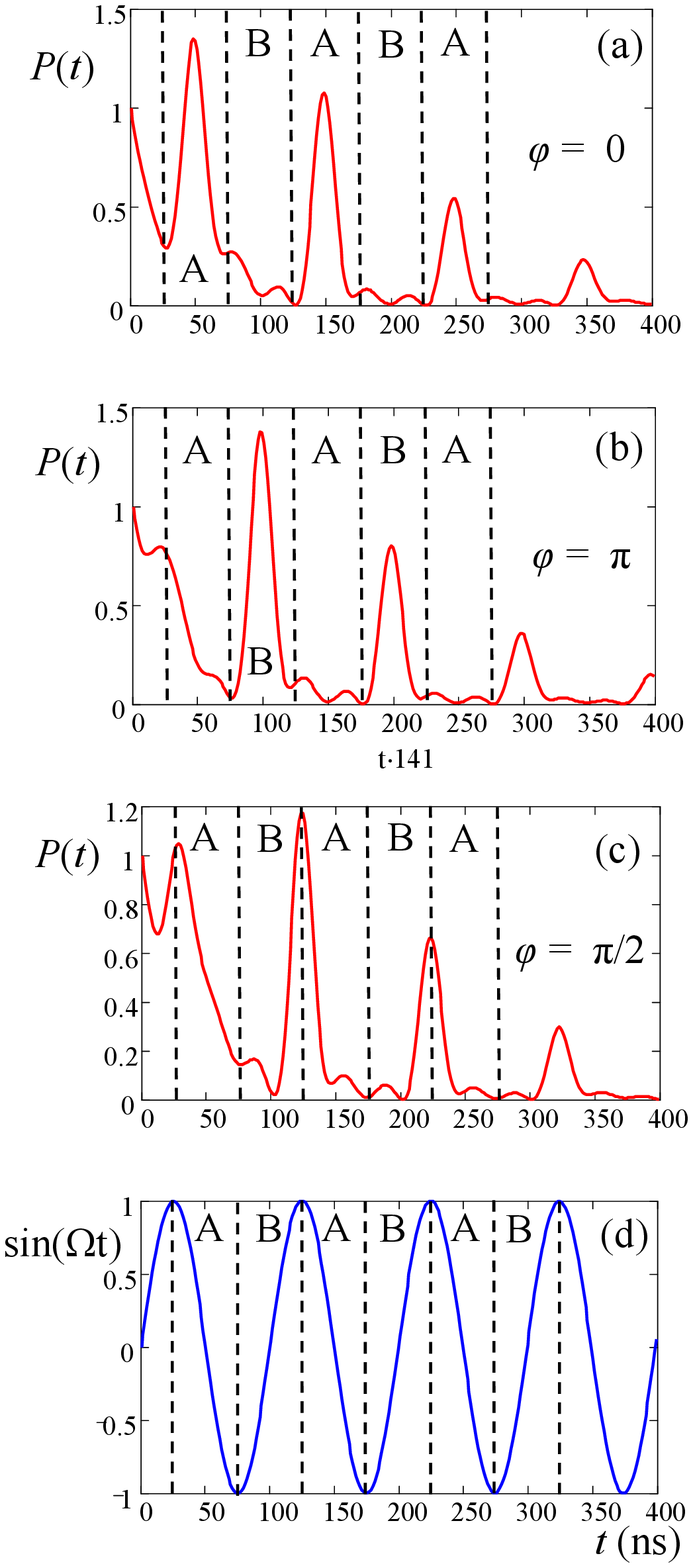}} \caption{(color on
line) (a)-(c) Time dependence of the detection probability of a photon, which
is in resonance with the first satellite of the central component of the
frequency comb, $\omega_{s}=\omega_{a}+\Omega$. The vibration frequency of the
absorber is $\Omega=10$MHz and the modulation index is $p=1.8$. Effective
thickness of the absorber is $T=12$. The value of the vibration phase
$\varphi$ is indicated in each plot. (d) The phase evolution of the field
interacting with the vibrating absorber in its reference frame. Dashed
vertical lines separate time bins A and B (see the text for details).}%
\label{fig:6}%
\end{figure}
\begin{figure}[ptb]
\resizebox{0.5\textwidth}{!}{\includegraphics{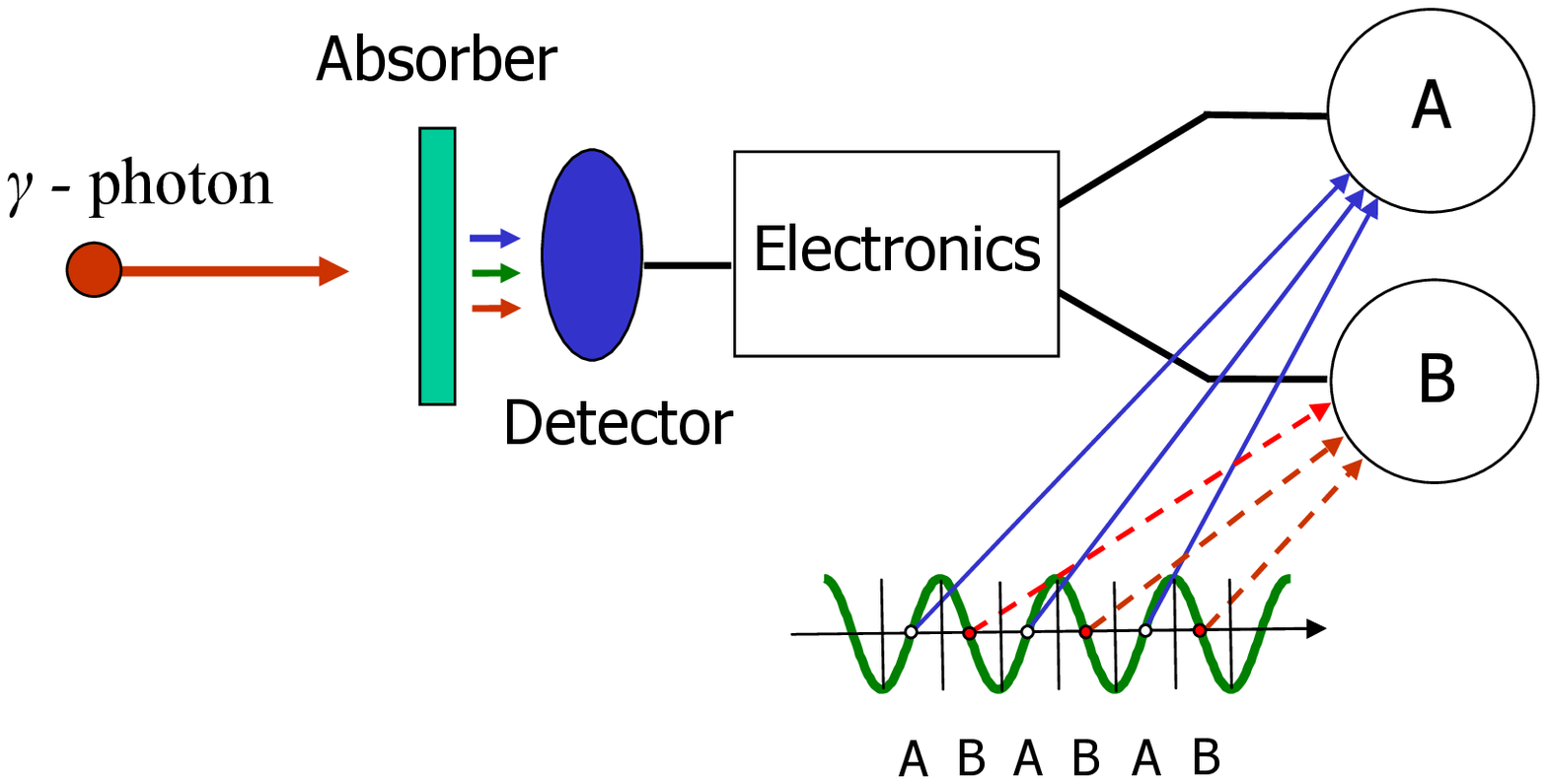}} \caption{(color on
line) Schematic presentation of the method how with one detector (dark oval in
blue) the pulses, formed by the vibrating absorber from single gamma-photon,
can be transferred by electronics (data acquisition system) into bins A and B.
The correspondence of these bins to the time evolution of the oscillating
phase (waving line in green) is shown in the bottom.}%
\label{fig:7}%
\end{figure}
\begin{figure}[ptb]
\resizebox{0.5\textwidth}{!}{\includegraphics{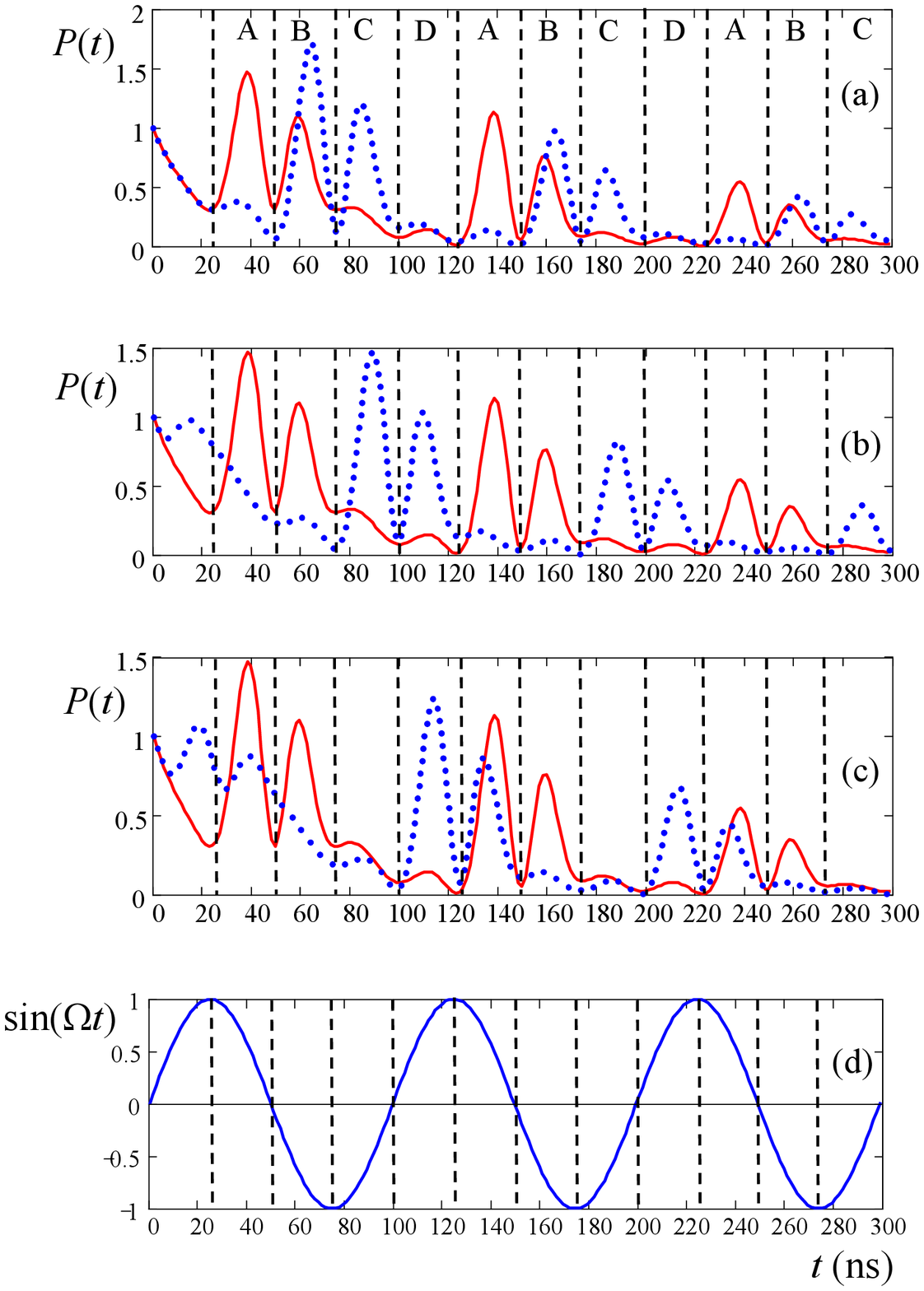}} \caption{(color on
line) (a)-(c) Time dependence of the detection probability of a photon with
comb spectrum, whose second satellite of the central component is in resonance
with the absorber, $\omega_{s}=\omega_{a}+2\Omega$. Frequency of the phase
modulation is $\Omega=10$MHz and the modulation index is $p=3.1$. Effective
thickness of the absorber is $T=12$. The value of the modulation phase
$\varphi$ is zero for solid line (in red) and it is $-\pi/2$ (a), $-\pi$ (b),
and $-3\pi/2$ (c) for dotted lines (in blue). (d) The phase evolution of the
field after the phase modulator (normalized to the modulation index $p$) if
$\varphi=\varphi_{lo}=0$. Dashed vertical lines separate time bins A, B, C,
and D (see the text for details).}%
\label{fig:8}%
\end{figure}

In optical domain such bins can be spatially separated by router based, for
example, on the Mach-Zehnder interferometer with a phase shifter placed in one
of the interferometer arms. If this phase shifter is fed by the local
oscillator, one can send the radiation field from bins A to the detector A and
from bins B to the detector B in accordance with the phase evolution shown in
Fig. 6d. If the phase modulator of the radiation field, placed between the
source and absorber to create a frequency comb, has the modulation phase
$\varphi$, which is the same as $\varphi_{lo}=0$, then only the detector A
will detect the radiation field. If this phase has a $\pi$ shift with respect
to $\varphi_{lo}=0$, only the detector B will detect the radiation field. If
the modulation phase is $\pi/2$, both detectors have the same probability of
photon detection. In such a way we propose to create time bin qubit, which is
equivalent to spin 1/2.

In gamma domain the routers are not currently available. However, main elements,
from which they could be constructed, are recently developed. They are
high-efficient back-reflecting mirrors \cite{Shvydko11}, efficient beam-splitters
\cite{Yamauchi}, tight focusing facilities \cite{Krebs}, and cavities \cite{Rentzepis}.
Meanwhile, even without routers we can distinguish A and B bins electronically
(see Fig. 7). In time delayed coincidence counting technique we have only two detectors.
One is for the heralding 122 keV photon, which starts the clock, and the other
is for the resonant 14.4 keV photon, which stops the clock. We distinguish
detection events of 14.4 keV photon in time by multichannel data acquisition
system with quite short duration of each channel (see, for example, Ref.
\cite{Shakhmuratov09,Shakhmuratov11}). This scheme can be easily modified to
simulate effective detectors A and B electronically, having physically only
one detector for the resonant gamma photon.

We assume that in optical domain it is possible to organize qubits of higher
dimension, known as qudits, by transformation of a single photon into bunches
of pulses and by routers. We estimate that it would be hard to vibrate the
absorber piston like with large amplitude comparable with the wavelength of
the optical radiation field. The simplest way to organize the phase modulation
of the radiation field with high frequency and large deviations is the use of
phase modulators. If the modulation index is large enough the single frequency
radiation field is transformed into a frequency comb with desirable
properties. As an example we consider the case when the second satellite of
the central component of the radiation field incident to the absorber is tuned
in resonance, $\omega_{s}=\omega_{a}+2\Omega$, and modulation index has
optimal value $p=3.1$, i.e., it is close to $\pi$. Then the single-photon wave
packet after passing through the absorber is split into bunches of pulses with
two pulses in each bunch (see Fig. 8a, solid curve in red). Now, time can be
grained into time bins A, B, C, and D with a duration $T_{v}/4$ each. If the
phase of the phase modulation (PM) is zero, $\varphi=0$, i.e. it coincides
with the phase of the local oscillator, $\varphi_{lo}$, then only bins A and B
contain the radiation pulses, while bins C and D fall into dark windows (see
Fig. 8a, solid curve in red). If the phase of PM is $\varphi=-\pi/2$, the
pulse bunches are shifted to a quarter of modulation period and then bins B
and C are occupied while bins A and D are almost empty (see Fig. 8a, dotted
line in blue). Changing the modulation phase further ($\varphi=-\pi$) we can
move pulses from bins B and C to bins C and D (see Fig. 8b, dotted line in
blue). If the modulation phase is $\varphi=-3\pi/2$, the pulses occupy D and A
bins only (see Fig. 8c, dotted line in blue).

To separate spatially time bins A, B, C, and D we propose to transmit the
radiation field through a set of routers R1, R2, and R3 (see Fig. 9). Router 1
(R1) separates the couple of time bins A and B from couple of time bins C and
D. R1 is synchronized with the local oscillator such that the first half a
period $T_{lo}$ of the local oscillator the radiation field is sent to the
router R2 and the second half of the oscillation period the radiation field is
sent to the Router 3. These routers, R2 and R3, switch the path of the
radiation field two times faster than R1 but with the same phase $\varphi
_{lo}$. Then the radiation field, contained in time bin A, will always go to
the detector A. The same is realized for time bins B, C, and D. The radiation
field contained in these time bins will go to the detectors B, C, and D,
respectively (see Fig. 9). By changing the phase of PM, $\varphi$, with
respect to the phase of local oscillator, $\varphi_{lo}$, one can control the
population of bins A, B, C, and D.
\begin{figure}[ptbptb]
\resizebox{0.5\textwidth}{!}{\includegraphics{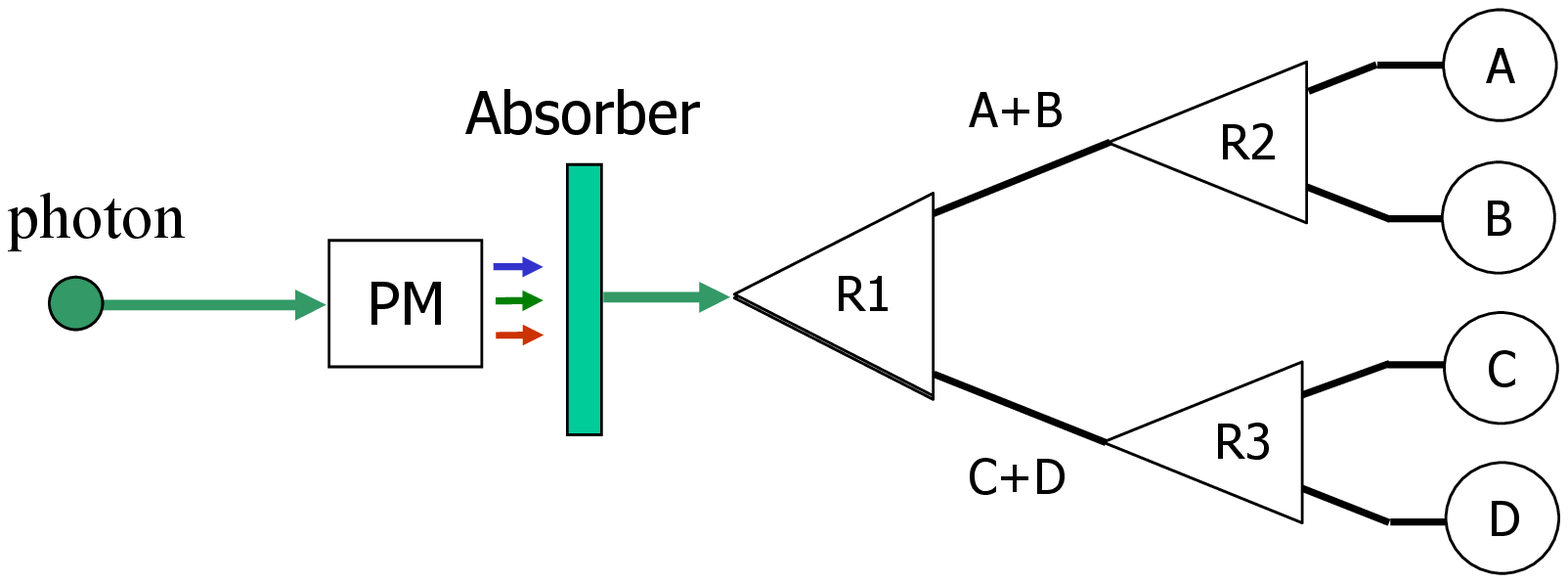}} \caption{(color on
line) Spatial separation of pulses from bunches. Single photon radiation field
is transformed to the frequency comb by phase modulator PM. Passing through
the absorber the single photon wave packet is shaped into pulses, which are
spatially separated by routers R1, R2, and R3 such that bins A, B, C, and D
are sent to the corresponding detectors A, B, C, and D (see the text for
details).}%
\label{fig:9}%
\end{figure}

It is interesting to notice that single pulses, generated when we tune in
resonance the first satellite (see Fig. 6), are in phase with the incident
radiation field. When we tune in resonance the second satellite, two pulses
are grouped in a bunch (see Fig. 8). The first pulse in a bunch has a phase
shift $\pi/2$ with respect to the incident field and the second pulse has
opposite phase, $-\pi/2$. This feature could be used to implement some
operations with bins A, B, C, and D if we make them interfere after passing
appropriate bins through a delay line.